\documentclass[a4paper,11pt]{article}

\usepackage{tikz}
\usepackage{graphicx}
\usepackage{multirow}
\usepackage{color}
\usepackage{xcolor}
\usepackage{latexsym}
\usepackage{amssymb}
\usepackage{amsfonts}
\usepackage{amsmath}
\usepackage{indentfirst}
\usepackage{slashbox,colortbl}
\usepackage[enableskew]{youngtab}
\usepackage{lscape}
\usepackage{epsfig}

\usetikzlibrary{matrix,fit}

\newcommand{\cvd}{\hfill $\blacksquare$\bigskip}
\newtheorem{definition}{Definition}[section]

\newtheorem{proposition}{Proposition}[section]

\date{}
\author{Elena Barcucci\thanks{Dipartimento di Matematica e Informatica ``U. Dini'', Universit\`a degli
Studi di Firenze, Viale
 G.B. Morgagni 65, 50134 Firenze, Italy. {
 \tt \ elena.barcucci@unifi.it,\quad antonio.bernini@unifi.it,\quad stefano.bilotta@unifi.it,\quad
renzo.pinzani@unifi.it}}\and Antonio Bernini$^*$ \and Stefano Bilotta$^*$ \and Renzo Pinzani$^*$}

\title{Non-overlapping matrices}

\begin{document}

\maketitle

\begin{abstract}
Two matrices are said non-overlapping if one of them can not be put on the other one in a way such that the corresponding entries coincide.
We provide a set of non-overlapping binary matrices and a formula to enumerate it which involves the $k$-generalized Fibonacci numbers. Moreover, the generating function for the enumerating  sequence is easily seen to be rational.
\end{abstract}

\section{Introduction}

A string $u$ over a finite alphabet $\Sigma$ is said \emph{self non-overlapping} (or equivalently \emph{unbordered} or \emph{bifix-free}) if it does not contain proper prefixes which are also proper suffixes. In other words, a string $u \in \Sigma^*$ is unbordered if it can not be factorized as $u=vu'v$ with $v \in \Sigma^+$ and $u' \in \Sigma^*$.
Nielsen in \cite{nielsen} provided the set $X \subset \Sigma^n$ of all bifix-free strings by means of a recursive construction. More recently, several researches \cite{bajic,bilo,black,singa} have been conducted in order to define particular subsets of $X$ constituted by \emph{non-overlapping} (or \emph{cross-bifix-free}) strings: two $n$ length strings $u,v \in X$ are said non-overlapping if any non-empty proper prefix of $u$ is different from any non-empty proper suffix of $v$, and viceversa.

In \cite{ans3} the notion of unbordered strings is generalized to the two dimensional case by means of \emph{unbordered pictures} which are rectangular matrices over $\Sigma$ by imposing that all possible overlaps between two copies of the same picture are forbidden. In particular, the authors extend in two dimensions the construction of unbordered strings proposed in \cite{nielsen} and describe an algorithm to generate the set $U$ of all the unbordered pictures of fixed size $m \times n$.

\begin{figure}[htb]
\begin{center}
\includegraphics[scale=0.5]{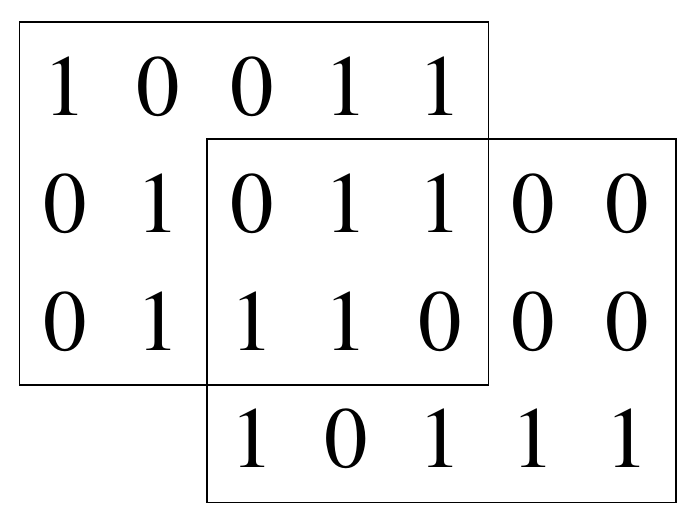} \caption{An example of overlap}
\end{center}\label{es1}
\end{figure}

The aim of the present paper is to find a subset of unbordered matrices which are non-overlapping. As well as the sets given in \cite{bajic,bilo,black,singa} are non-overlapping subsets (or cross-bifix-free subsets) of strings of $X$, in the same way the set we are going to present is a non-overlapping subset of matrices of $U$. Roughly speaking two unbordered matrices $A$ and $B$ are non-overlapping if all possible overlaps between $A$ and $B$ are forbidden. More precisely, we can imagine to make a rigid movement of $B$ on $A$ such that $B$ glides on $A$. At the end of each slipping, which can be geometrically interpreted as a translation in a given direction on the plane, a (non empty) common area (in the sequel \emph{control window}) is formed. This common area can be seen as the usual intersection between the two rectangular arrays containing the entries of $A$ and $B$, which is, in turn, a rectangular array constituted by a finite number of $1\times 1$ cells of the discrete plane. Each cell of the control window contains an entry of $A$ and an entry of $B$. If in each cell of the window the entry of $A$ coincides with the entry of $B$, then such window is said \emph{overlapping window} and $A$ and $B$ \emph{overlapping matrices}. On the contrary, if for any translation we never find an overlapping window, $A$ and $B$ are said \emph{non-overlapping matrices}.  For example, the unbordered matrices
$A=\begin{pmatrix}
1&0&0&1&1\\
0&1&0&1&1\\
0&1&1&1&0\\
\end{pmatrix}$
and
$
B=
\begin{pmatrix}
0&1&1&0&0\\
1&1&0&0&0\\
1&0&1&1&1\\
\end{pmatrix}
$
can be overlapped as in Figure \ref{es1} where the control window is showed.

%the next diagram where the bold characters shows the control window:
%
%$$\begin{matrix}
%1 & 0 & 0 & 1 & 1 &   &  \\
%0 & 1 & \bf{0} & \bf{1} & \bf{1} & 1 & 0\\
%0 & 1 & \bf{1} & \bf{1} & \bf{0} & 0 & 0\\
%  &   & 0 & 0 & 1 & 0 & 0\\
%\end{matrix}
%$$

Actually, a first attempt in order to generalize the concept of non-overlap in two dimensions between two distinct matrices can be found in \cite{tcs} where the authors define a set of \emph{cross-bibifix-free} square matrices over a finite alphabet. For the sake of clearness, two square matrices are said to be cross-bibifix-free when, essentially, they are non-overlapping only along the direction of the main diagonal. Here, using a completely different approach, we consider translations in any direction on the plane and matrices which can be also rectangular matrices, even if they have only binary entries. In this way the definition of non-overlapping set of matrices we are going to propose seems to be very close to the natural generalization in two dimensions of the concept of non-overlapping set of strings.

As it often happens, the
extension to the bidimensional case of a typical concept
related to strings is carried on by taking into account matrices.
There are several cases in the literature where this process is occurred. For example, in \cite{croc} a
bidimensional variant of the string matching problem is considered for sets of matrices. Another interesting example is given by the
extension of classical finite automata for strings to the two-dimensional rational automata for pictures introduced in \cite{ans}. Moreover, it is worth to mention the problem of the pattern avoidance in matrices \cite{kita}, which is a typical topic in linear structures as permutations and words.

\bigskip

In Section \ref{set_nonov} we formally define a set of binary matrices which are proved to be non-overlapping matrices. The cardinality of this set is given in Section \ref{number} where we also show that it is related to the well-known $k$-generalized Fibonacci numbers.

\section{A set of non-overlapping binary matrices}\label{set_nonov}

The definition of non-overlapping matrices given in the Introduction can be formalized in terms of blocks matrices. Indeed, the control window we have referred in the previous section is essentially a particular block whose dimensions impose the ones of the other blocks of the partition of the matrix.

\begin{definition}
Let $\mathcal M_{m\times n}$ be the set of all the matrices with $m$ rows and $n$ columns. Two distinct matrices $A,B \in \mathcal M_{m\times n}$ are said \emph{non-overlapping} if all the following conditions are satisfied by $A$ and $B$:
\begin{itemize}

\item there do not exist two block partitions
$$A = \begin{bmatrix} A_{11} & A_{12}\\ A_{21} & A_{22}\end{bmatrix} \mbox{and}\  B = \begin{bmatrix} B_{11} & B_{12}\\ B_{21} & B_{22}\end{bmatrix}$$ such that $A_{11},B_{22}\in \mathcal M_{r\times s}$, with $1\leq r \leq m-1$, $1\leq s \leq n-1$, and neither $A_{11}=B_{22}$, nor $A_{12}=B_{21}$, nor $A_{21}=B_{12}$, nor $A_{22}=B_{11}$.

\item there do not exist two block partitions
$$A= \begin{bmatrix} A_{11} \\ A_{21}\end{bmatrix} \mbox{and}\  B=\begin{bmatrix} B_{11} \\ B_{21}\end{bmatrix}$$ such that $A_{11},B_{21}\in \mathcal M_{r\times n}$, with $1\leq r\leq m-1$, and neither $A_{11}=B_{21}$, nor $A_{21}=B_{11}$.

\item there do not exist two block partitions
$$A= \begin{bmatrix} A_{11} & A_{12}\end{bmatrix} \mbox{and}\ B=\begin{bmatrix} B_{11} & B_{12}\end{bmatrix}$$ such that $A_{11},B_{12}\in \mathcal M_{m\times s}$, with $1\leq s\leq n-1$, and neither $A_{11}=B_{12}$, nor $A_{12}=B_{11}$.
\end{itemize}

\end{definition}

In other words, two distinct matrices are non-overlapping if any control window is not an overlapping window. Therefore, we can also define a \emph{self non-overlapping} (or \emph{unbordered}) matrix $A\in \mathcal M_{m\times n}$ as a matrix such that there does not exist a translation of $A$ on itself such that we never find an overlapping window. Clearly, this last definition can be easily deduced from Definition 2.1 with $A=B$ and suitably adapting the block partitions.

%
%\begin{itemize}
%
%\item there not exists a block partition
%$$A = \begin{bmatrix}
%A_{11} & A_{12} & A_{13}\\
%A_{21} & A_{22} & A_{23}\\
%A_{31} & A_{32} & A_{33}
%\end{bmatrix}
%$$
%such that $A_{11},A_{33}\in \mathcal M_{r\times s}$, with $1\leq r \leq m-1$, $1\leq s \leq n-1$, and neither $A_{11}=A_{33}$, nor $A_{13}=A_{31}$.
%
%
%\item there not exists a block partition
%$$A= \begin{bmatrix} A_{11}\\
%A_{21}\\
%A_{31}
%\end{bmatrix}
%$$
%such that $A_{11},A_{31}\in \mathcal M_{r\times n}$, with $1\leq r\leq m-1$, and $A_{11}=A_{31}$.
%
%\item there not exists a block partition
%$$A= \begin{bmatrix}
%A_{11} & A_{12}& A_{13}
%\end{bmatrix}
%$$
%such that $A_{11},A_{13}\in \mathcal M_{m\times s}$, with $1\leq s\leq n-1$, and $A_{11}=A_{13}$.
%\end{itemize}
%
%\end{definition}

%Clearly, in the above definition, if $A=B$, opportunely adapting the block partitions of $A$, then the matrix does not overlap with itself and in this case it is said \emph{self non-overlapping}.

%In the case of square matrix $A$, following \ref{croc}, a square block is a \emph{border} if it occurs at the four corners

\begin{definition}

A set $\mathcal{S}_{m\times n} \subset \mathcal{M}_{m\times n}$ is called \emph{non-overlapping} if each matrix of $\mathcal S_{m\times n}$ is self non-overlapping and for any two matrices $A,B\in \mathcal S_{m\times n}$ they are non-overlapping matrices.

\end{definition}

Fixed the dimension $m\times n$ of the matrices, we now define a possible non-overlapping set where the matrices have a particular structure involving some of the entries on the frame of the matrix.

\begin{definition}\label{matrix}
Let $3\leq k\leq \left\lfloor\frac{n}{2}\right\rfloor$. We denote $\mathcal S^{(k)}_{m\times n}\subset \mathcal M_{m\times n}$ the set of the matrices $A=\left(a_{i,j}\right)$ satisfying the following conditions:

\begin{itemize}

\item $A_1=1^{k-1}0w_110^{k-1}$, where $v_1=0w_11$ is a binary string of length $n-2k+2$ avoiding both $0^k$ and $1^k$;

\item for $i=2,\ldots,m-1$, $A_i=w_i0=v_i$, where $v_i$ is a binary string of length $n$ avoiding both $0^k$ and $1^k$;

\item $A_m=1^kv_m0^k$, where $v_m$ is a binary string of length $n-2k$ avoiding both $0^k$ and $1^k$.
		
(With $A_1$, $A_i$ and $A_m$ we denote the first, the $i$-th and the $m$-th row of the matrix $A$.)		
		
\end{itemize}
\end{definition}

In other words, some entries on the frame of a matrix in $\mathcal S^{(k)}_{m\times n}$ are fixed. For example, the matrices in $\mathcal S^{(3)}_{6\times 10}$ are represented in Figure \ref{set} where the generic entries $*\in\{0,1\}$ are chosen so that the conditions of Definition \ref{matrix} are satisfied.
We note that for $k=2$ and $n$ odd the set $\mathcal S^{(2)}_{m\times n}$ can not be defined since the strings $v_i$ can not avoid both $00$ and $11$.

\begin{figure}
$$
\begin{pmatrix}
\bf{1}&\bf{1}&\bf{0}&*&*&*&*&\bf{1}&\bf{0}&\bf{0}\\
*&*&*&*&*&*&*&*&*&\bf{0}\\
*&*&*&*&*&*&*&*&*&\bf{0}\\
*&*&*&*&*&*&*&*&*&\bf{0}\\
*&*&*&*&*&*&*&*&*&\bf{0}\\
\bf{1}&\bf{1}&\bf{1}&*&*&*&*&\bf{0}&\bf{0}&\bf{0}\\
\end{pmatrix}
$$
\caption{The structure of the matrices in $\mathcal S^{(3)}_{6\times 10}$.}
\label{set}
\end{figure}

\begin{proposition}\label{prop1} The set $\mathcal S^{(k)}_{m\times n}\subset \mathcal M_{m\times n}$ is non-overlapping, for each $k$ with $3\leq k\leq \left\lfloor\frac{n}{2}\right\rfloor$, $m \geq 2$ and $n \geq 2k$.
\end{proposition}

\begin{figure}[htb]
\begin{center}
\includegraphics[scale=0.5]{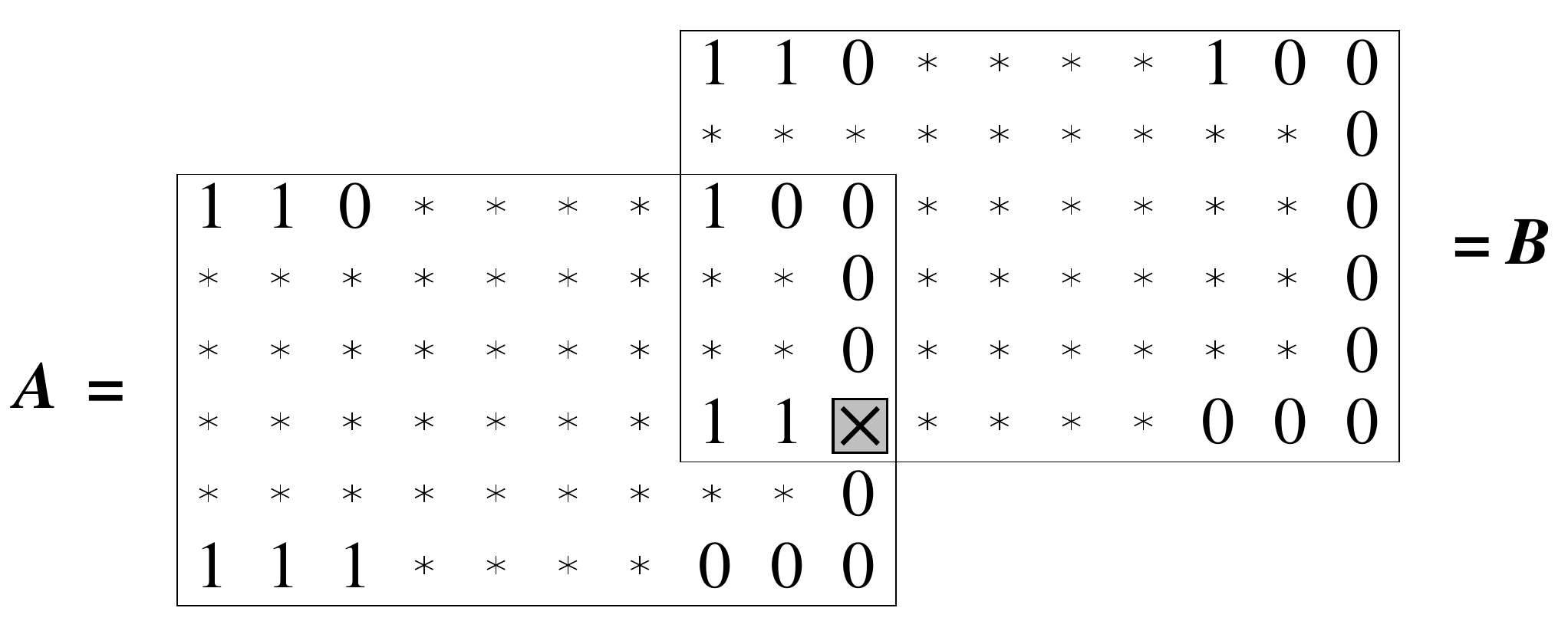} \caption{The slipping of case 2: the grey entry contains different values of $A$ and $B$.}\label{proof2}
\end{center}
\end{figure}

\emph{Proof.} \quad Given to matrices $A, B \in \mathcal S^{(k)}_{m\times n}$, we present two possible slippings of $B$ on $A$ (without loss of generality $A$ and $B$ can be interchanged).
\begin{enumerate}

\item In the case represented in Figure \ref{proof1} the obtained control window contains, in its lower row, a string having $k$ consecutive equal symbols. If $A$ and $B$ would be overlapping matrices, then these $k$ consecutive equal symbols, belonging to the frame of $B$, should also appear in a row $v_i$ of $A$, against the hypothesis that $A \in \mathcal S^{(k)}_{m\times n}$.

\begin{figure}[htb]
\begin{center}
\includegraphics[scale=0.5]{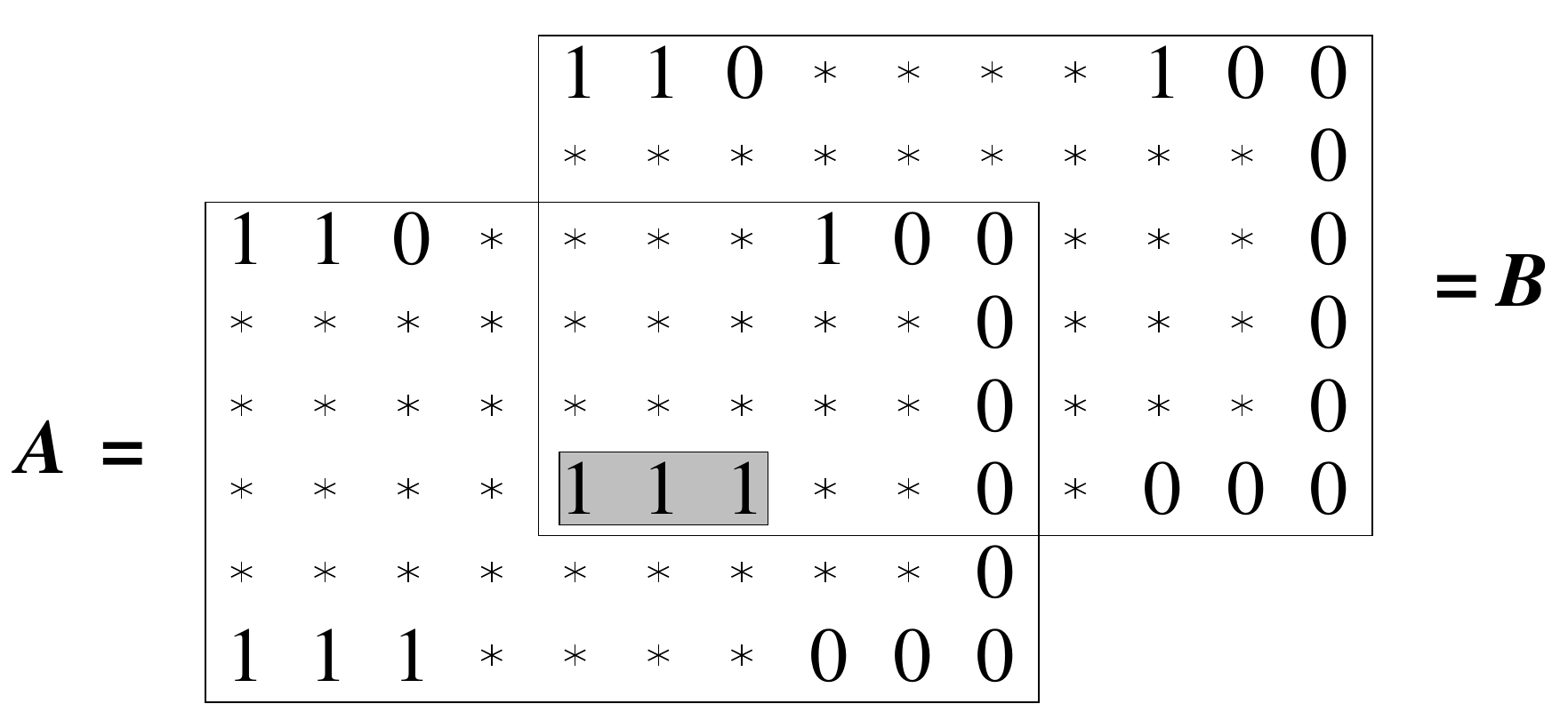} \caption{The slipping of case 1: the grey entries are a forbidden sequence of $A$.}\label{proof1}
\end{center}
\end{figure}

%
%\begin{figure}
%$$
%A=
%\begin{array}{cccccccccccccc}
%&&&&1&1&0&*&*&*&*&1&0&0\\
%&&&&1&*&*&*&*&*&*&*&*&0\\
%1&1&0&*&1&*&*&1&0&0&*&*&*&0\\
%1&*&*&*&1&*&*&*&*&0&*&*&*&0\\
%1&*&*&*&1&*&*&*&*&0&*&*&*&0\\
%1&*&*&*&\colorbox{black!20}1&\colorbox{black!20}1&\colorbox{black!20}1&*&*&0&*&0&0&0\\
%1&*&*&*&*&*&*&*&*&0&&&&\\
%1&1&1&*&*&*&*&0&0&0&&&&\\
%\end{array}
%=B
%$$
%\caption{The slipping of case 1: the gray entries are a forbidden sequence of $A$.}
%\label{proof1}
%\end{figure}

\item Another possible slipping is pictured in Figure \ref{proof2} where the control window is non-overlapping (and then $A$ and $B$ are non-overlapping matrices) since it presents certain cells where the fixed entries of $A$ do not coincide with the fixed entries of $B$. These fixed entries belong to the frame of $A$ and $B$.
\end{enumerate}

%
%\begin{figure}
%$$
%A=
%\begin{array}{ccccccccccccccccc}
%&&&&&&&1&1&0&*&*&*&*&1&0&0\\
%&&&&&&&1&*&*&*&*&*&*&*&*&0\\
%1&1&0&*&*&*&*&1&0&0&*&*&*&*&*&*&0\\
%1&*&*&*&*&*&*&1&*&0&*&*&*&*&*&*&0\\
%1&*&*&*&*&*&*&1&*&0&*&*&*&*&*&*&0\\
%1&*&*&*&*&*&*&1&1&\colorbox{black!20}{$\times$}&*&*&*&*&0&0&0\\
%1&*&*&*&*&*&*&*&*&0&&&&&&&\\
%1&1&1&*&*&*&*&0&0&0&&&&&&&\\
%\end{array}
%=B
%$$
%\caption{The slipping of case 2: the gray entry contains different values of $A$ and $B$.}
%\label{proof2}
%\end{figure}

In general, given a control window of dimension $r\times s$,  with $1\leq r\leq m$ and $1\leq s\leq n$, we can refer to one of the two above cases depending on the values of $r$ and $s$. In particular we have:

\begin{itemize}
\item $r=1:$
\begin{itemize}
\item if $1\leq s\leq2k-2$, there exists a slipping as in case 2;
\item if $2k-1\leq s\leq n-k$, there exists a slipping as in case 1;
\item if $n-k+1\leq s\leq n$, there exists a slipping as in case 2;

\end{itemize}
\item $2\leq r\leq m-1:$
\begin{itemize}
\item if $1\leq s\leq k$, there exists a slipping as in case 2;
\item if $k+1\leq s\leq n$, there exists a slipping as in case 1;

\end{itemize}

\item $r=m:$
\begin{itemize}
\item if $1\leq s\leq 2k-1$, there exists a slipping as in case 2;
\item if $2k\leq s\leq n-k$, there exists a slipping as in case 1;
\item if $n-k+1\leq s \leq n-1$, there exists a slipping as in case 2.
\end{itemize}

\end{itemize}

To complete the proof, if $A=B$, then it immediately follows from the above argument that the matrices of $\mathcal S^{(k)}_{m\times n}$ are self non-overlapping.

\cvd

%
%
%\begin{equation*}
%    U^{k} = \begin{tikzpicture}
%            \matrix [matrix of math nodes,left delimiter=(,right delimiter=)] (m)
%            {
%                0     &  0 & \cdots & s z^{0} & s z^{1} & \cdots & s z^{k-2} & s    z^{k-1} \\
%                0  &  \vdots&    & -\overline{z} & s^{2} z^{0}   & \cdots & s^{2} z^{k-3} & s^{2} z^{k-2} \\
%                0 &   & &  & -\overline{z} & s^{2} z^{0} & \cdots & s^{2} z^{k-3} \\
%                \vdots    &   & &  &   & \ddots & \ddots & \vdots \\
%                z s       & \vdots&  &   & 0 &   & -\overline{z} & s^{2} z^{0} \\
%                s         & 0 & \cdots &  &   &   &   & -\overline{z} \\
%                0         & 1 &   & 0 &  &  &   &   \\
%                \vdots    &   & \ddots & &  &   & 0 &   \\
%                0         & 0 &   & 1 &  & &   &   \\
%            };
%
%            \draw (m-7-2.north west) arc (10:-10:-12);
%            \end{tikzpicture}
%\end{equation*}

\section{The enumeration of $\mathcal S_{m\times n}^{(k)}$}\label{number}
In this section we are going to enumerate the set  $\mathcal S^{(k)}_{m\times n}$. It is easy to realize that its cardinality depends on the number of rows satisfying the constraints of Definition \ref{matrix}.

We denote by $R_n(0^k,1^k)$ be the set of binary string starting with $0$, ending with $1$ and avoiding $k$ consecutive $0$'s and $k$ consecutive $1$'s. Let $Z_n(0^k,1^k)$ be the set of binary strings ending with $0$ and avoiding $k$ consecutive $0$'s and $k$ consecutive $1$'s.
Moreover, let $B_n(0^k,1^k)$ the set of binary strings avoiding $k$ consecutive $0$'s and $k$ consecutive $1$'s. We indicate with $r_n^{(k)}$, $z_n^{(k)}$ and  $b_n^{(k)}$ the
cardinality of $R_n(0^k,1^k)$, $Z_n(0^k,1^k)$ and $B_n(0^k,1^k)$,
respectively. It is straightforward that

\begin{equation}\label{totmatrix}
|\mathcal S^{(k)}_{m\times n}|=r_{n-2k+2}^{(k)}\cdot \left(z_n^{(k)}\right)^{m-2}\cdot b_{n-2k}^{(k)}
\end{equation}

where, referring to Definition \ref{matrix}, the term $r_{n-2k+2}^{(k)}$ counts the number of strings $v_1$, the terms  $z_n^{(k)}$ count the number of strings $v_i$ for $i=2,3,\ldots,m-1$, and $b_{n-2k}^{(k)}$ is the number of strings $v_m$.

\subsection{The sequence $r_n^{(k)}$}

Now we consider a possible recursive relation for $r_n^{(k)}$ by means of a recursive construction of $R_n(0^k,1^k)$. We first observe that $R_0(0^k,1^k) =\{\lambda\}$, $R_1(0^k,1^k)=\emptyset$ and $R_j(0^k,1^k)$ is formed by all the binary strings of length $j$, with $2 \leq j \leq k$, starting with $0$ and ending with $1$. Then $r_0^{(k)}=1$, $r_1^{(k)}=0$ and $r_j^{(k)}=2^{j-2}$ for $2 \leq j \leq k$. Clearly, if $k=2$, then $r_n^{(k)}=0$ in the case of $n$ odd and $r_n^{(k)}=1$ if $n$ is even (in this case $R_n(0^k,1^k)=\{\underbrace{0101\ldots01}_{n}\}$).

Fixed $k \geq 3$ and $n\geq k+1$, each string $u \in R_n(0^k,1^k)$ can be factorized as $u=u' 0^i1^j$, with $1 \leq i,j \leq k-1$, and $u' \in R_{n-i-j}(0^k,1^k)$. Denoting with $h=i+j$ the length of the suffix $0^i1^j$, it is $2 \leq h \leq 2k-2$.

If $2 \leq h \leq k$, then $i$ can assume the values $1,2,\ldots,h-1$ and consequently $j=h-1,h-2,\ldots,1$ for a total of $h-1$ possibilities for the suffix $0^i1^j$, for each fixed $h$. Indeed, in this case, the suffix $0^i1^j$ contains neither $0^k$ nor $1^k$.

If $k+1 \leq h \leq 2k-2$, in order to avoid the forbidden patterns $i$ can assume the values $h-k+1, h-k+2, \ldots, k-1$ and consequently $j=k-1,k-2,\ldots,h-k+1$ for a total of $2k-h-1$ possibilities for the suffix $0^i1^j$, for each fixed $h$.

Therefore, for $n \geq k+1$,

$$r_n^{(k)} = \sum_{h=2}^{k} (h-1)r_{n-h}^{(k)}+ \sum_{h=k+1}^{2k-2}(2k-h-1)r_{n-h}^{(k)}.$$

Summarizing:
\begin{equation*}
r_n^{(k)}=\left \{
\begin{array}{ll}
1 & \mbox{if} \ n =0\\
\\
0 & \mbox{if} \ n =1\\
\\
2^{n-2} & \mbox{if} \ 2\leq n \leq k\\
\\
\displaystyle\sum_{h=2}^{k} (h-1)r_{n-h}^{(k)}+ \displaystyle\sum_{h=k+1}^{2k-2}(2k-h-1)r_{n-h}^{(k)} & \mbox{if} \  n \geq k+1.\\
\end{array}
\right.
\end{equation*}

Note that the coefficients of $r_n^{(k)}$, for $n\geq k+1$, are the coefficients of Smarandache Crescendo Pyramidal sequence (see sequence A004737 in The On-line Encyclopedia of Integer Sequence).
In Table \ref{rn} we list the first numbers of the recurrence $r_n^{(k)}$ for some fixed values of $k$.

\begin{table}
\begin{center}
{\footnotesize
\begin{tabular}{|c|rrrrrr|}
\hline
\backslashbox{$n$}{$k$} & 3 & 4 & 5 & 6 & 7 & 8\\
\hline
0 & 1 & 1 & 1 & 1 & 1 & 1\\
1 & 0 & 0 & 0 & 0&0&0\\
2 & 1 & 1 & 1 & 1&1&1\\
3 & 2 & 2 & 2 & 2&2&2\\
4 & 2 & 4 & 4 & 4&4&4 \\
5 & 4 & 6 &  8&8&8&8  \\
6 & 7 & 12 & 14 & 16 & 16 & 16\\
7&10&22&28&30&32&32\\
8&17&41&54&60&62&64\\
9&28&74&104&118&124&126\\
10&44&137&201&232&246&252\\
11&72&252&386&456&488&502\\
12&117&464&745&897&968&1000\\
13&188&852&1436&1762&1920&1992\\
14&305&1568&2768&3465&3809&3968\\
15&494&2884&5336&6812&7554&7904\\

\hline\end{tabular}
}
\caption{Sequences $r_n^{(k)}$ for some fixed values of $k$.} \label{rn}
\end{center}
\end{table}

\bigskip

Since the strings $u\in R_n(0^k,1^k)$ have been factorized with $u=u'0^i1^j$, where the suffix $0^i1^j$ has length at least 2, the first term in the recurrence for $r_n^{(k)}$ is $r_{n-2}^{(k)}$. In the following we provide another construction for the strings $u$ leading to a recurrence for $r_n^{(k)}$ which involves also the term $r_{n-1}^{(k)}$.

\medskip
\noindent
Let $R_n^{(1^i)}(0^k,1^k)$ be the subset of $R_n(0^k,1^k)$ of the strings ending with $i$ ones, with $i=1,2,\ldots,k-1$. Then $R_n(0^k,1^k)=\displaystyle\bigcup_{i=1}^{k-1}R_n^{(1^i)}(0^k,1^k)$.

The strings of $R_n^{(1^1)}(0^k,1^k)$ are obtained by the strings of $R_{n-j-1}(0^k,1^k)$ appending $0^j1$, for $j=1,2,\ldots,k-1$. So that
\begin{equation}\label{r1}
\left| R_n^{(1^1)}(0^k,1^k) \right|=r_{n-2}^{(k)}+r_{n-3}^{(k)}+\ldots +r_{n-k}^{(k)}\ .
\end{equation}

The strings of $R_n^{(1^i)}(0^k,1^k)$, with $2\leq i\leq k-1$, can be obtained from all the strings of $R_{n-1}(0^k,1^k)$, appending one $1$ at the end of each string, excluding the strings in $R_{n-1}^{(1^{k-1})}(0^k,1^k)$ since the forbidden pattern $1^k$ would appear. So,
\begin{equation}\label{ri}
\sum_{i=2}^{k-1}\left| R_n^{(1^i)}(0^k,1^k) \right|=r_{n-1}^{(k)}-
\left| R_{n-1}^{(1^{k-1})}(0^k,1^k) \right|\ .
\end{equation}

For what the last term is concerned, we observe that the strings of $R_{n-1}^{(1^{k-1})}(0^k,1^k)$ are obtained from the strings of $R_{n-1-(k-2)}^{(1^1)}(0^k,1^k)$ appending $1^{k-2}$. Clearly, $\left| R_{n-1}^{(1^{k-1})}(0^k,1^k) \right|=\left| R_{n-k+1}^{(1^1)}(0^k,1^k) \right|$. The set
$R_{n-k+1}^{(1^1)}(0^k,1^k)$ is obtained from the strings of $R_{n-k+1-(j+1)}(0^k,1^k)$ appending $0^j1$, for $j=1,2,\ldots,k-1$. Then,
\begin{equation}\label{rk}
\left| R_{n-1}^{(1^{k-1})}(0^k,1^k) \right|=
r_{n-k-1}^{(k)}+r_{n-k-2}^{(k)}+\ldots+r_{n-2k+1}^{(k)}\ .
\end{equation}

Summarizing, since
$$r_n^{(k)}= \left| R_n^{(1^1)}(0^k,1^k) \right|
+\sum_{i=2}^{k-1}\left| R_n^{(1^i)}(0^k,1^k) \right|\ ,
$$
from Equation (\ref{r1}), (\ref{ri}) and (\ref{rk}), we obtain:
\begin{equation}\label{rn_meni}
r_n^{(k)}=\sum_{j=1}^{k}r_{n-j}^{(k)}-\sum_{j=k+1}^{2k-1}r_{n-
j}^{(k)}\ ,
\end{equation}
with initial conditions $r_{-i}^{(k)}=0$, $r_0^{(k)}=1$,
$r_1^{(k)}=0$.

%
%
%r_n^{(k)}=r_{n-1}^{(k)}+r_{n-2}^{(k)}+\ldots+r_{n-k}^{(k)}-
%r_{n-k-1}^{(k)}-r_{n-k-2}^{(k)}-\ldots-r_{n-2k+1}^{(k)}\ ,

We observe that the sequence
$r_n^{(k)}$ satisfies also a recurrence which is used in the rest of the paragraph. The $n$-th term of such a recurrence is given by the sum of the $k-1$ preceding terms plus 0 or 1 or -1, depending on $n$ and $k$. The result is stated in the following proposition.

\begin{proposition}\label{rn_positive}
Let $r_{-i}^{(k)}=0$, $r_0^{(k)}=1$,
$r_1^{(k)}=0$ be the initial conditions, then
\begin{equation*}
r_n^{(k)}=
\displaystyle\sum_{j=1}^{k-1} r_{n-j}^{(k)}+ d^{(k)}_n \ ,
\end{equation*}
where
\begin{equation*}
d^{(k)}_n=
\left\{
\begin{array}{rl}
1& \mbox{if}\ (n\ \mbox{mod}\ k)=0\\
\\
-1& \mbox{if}\ (n\ \mbox{mod}\ k)=1\\
\\
0& \mbox{if}\ (n\ \mbox{mod}\ k)\geq 2\ .\\
\end{array}
\right.
\end{equation*}

\end{proposition}

\emph{Proof.} \quad We can proceed by induction. If $n=2$, we have that $r_2^{(k)} = r_1^{(k)} + r_0^{(k)} + r_{-1}^{(k)} + \ldots + r_{3-k}^{(k)} + d_2^{(k)}$. Since $k\geq 3$ and $r_{-1}^{(k)}=0$, then $d_2^{(k)}=0$ and $r_2^{(k)} = r_1^{(k)} + r_0^{(k)}=1$ which is the same value obtained by recurrence (\ref{rn_meni}). Suppose that $r_s^{(k)}=\sum_{j=1}^{k-1} r_{s-j}^{(k)}+ d^{(k)}_s$ for each $s<n$. We have, from recurrence (\ref{rn_meni}),
$$
r_n^{(k)}=r_{n-1}^{(k)}+ \ldots +r_{n-k+1}^{(k)}+r_{n-k}^{(k)}-\sum_{k+1}^{2k-1}r_{n-
j}^{(k)}\ .
$$
For the inductive hypothesis
$$r_{n-k}^{(k)}= r_{n-k-1}^{(k)}+ \ldots + r_{n-2k+1}^{(k)} + d_{n-k}^{(k)} \ .$$
Then,
$$
r_n^{(k)}=r_{n-1}^{(k)}+ \ldots +r_{n-k-1}^{(k)}+d_{n-k}^{(k)}\ .
$$

Since $d_{n-k}^{(k)}=d_n^{(k)}$, the thesis follows. \cvd

\subsection{The sequences $b_n^{(k)}$ and $z_n^{(k)}$}

We give a recursive relation for $b_n^{(k)}$ by means of a recursive construction of the set $B_n(0^k,1^k)$. We first observe that $B_0(0^k,1^k) =\{\lambda\}$, where $\lambda$ is the empty string, and $B_j(0^k,1^k)$ is formed by all the binary strings of length $j$, with $0<j<k$. Then $b_j^{(k)}=2^j$ for $0 \leq j < k$.

In order to consider the strings in $B_n(0^k,1^k)$ having length $n \geq k$, we denote by $B_n^{(0)}(0^k,1^k)$ and $B_n^{(1)}(0^k,1^k)$ the two subsets of $B_n(0^k,1^k)$ constituted by the strings ending with $0$ and ending with $1$, respectively.
Let $|B_n^{(0)}(0^k,1^k)|=b_{n,0}^{(k)}$ and $|B_n^{(1)}(0^k,1^k)|
=b_{n,1}^{(k)}$, it is easy to realize that
$b_{n,0}^{(k)}=b_{n,1}^{(k)}= b_n^{(k)}/2$.

The set $B_n^{(0)}(0^k,1^k)$ can be generated from the strings in $B_{n-j}^{(1)}(0^k,1^k)$ followed by the suffix $0^j$, with $0 < j < k$, so we have
$$b_{n,0}^{(k)}=\sum_{j=1}^{k-1} b_{n-j,1}^{(k)}, \ \mbox{for} \ n \geq k.$$

Analogously, the set $B_n^{(1)}(0^k,1^k)$ can be generated from the strings in $B_{n-j}^{(0)}(0^k,1^k)$ followed by the suffix $1^j$, with $0 < j < k$, so we have
$$b_{n,1}^{(k)}=\sum_{j=1}^{k-1} b_{n-j,0}^{(k)}, \ \mbox{for} \ n \geq k.$$

Therefore, for $n \geq k$,
$$b_n^{(k)} = b_{n,0}^{(k)} + b_{n,1}^{(k)}=\sum_{j=1}^{k-1} \left( b_{n-j,1}^{(k)} + b_{n-j,0}^{(k)} \right)= \sum_{j=1}^{k-1} b_{n-j}^{(k)}.$$

Summarizing:
\begin{equation}\label{nobordi}
b_n^{(k)}=\left \{
\begin{array}{ll}
2^n & \mbox{if} \ 0 \leq n \leq k-1\\
\\
b_{n-1}^{(k)}+b_{n-2}^{(k)}+\ldots+b_{n-k+1}^{(k)} & \mbox{if} \ n \geq k.
\end{array}
\right.
\end{equation}

In Table \ref{bn} we list the first numbers of the recurrence $b_n^{(k)}$ for same fixed values of $k$.

\begin{table}
\begin{center}
{\footnotesize
\begin{tabular}{|c|rrrrrr|}
\hline
\backslashbox{$n$}{$k$} & 3 & 4 & 5 & 6 & 7 & 8\\
\hline
0 & 1 & 1 & 1 & 1 & 1 & 1\\
1 & 2 & 2 & 2 & 2 & 2 & 2\\
2 & 4 & 4 & 4 & 4 & 4 & 4\\
3 & 6 & 8 & 8 & 8 & 8 & 8\\
4 & 10 & 14 & 16 & 16 & 16 & 16\\
5 & 16 & 26 &  30& 32 & 32 & 32\\
6 & 26 & 48 & 58 & 62 & 64 & 64\\
7 & 42 & 88 & 112 & 122 & 126 & 128\\
8&68&162&216&240&250&254\\
9&110&298&416&472&496 & 506\\
10&178&548&802&928& 984 & 1008\\
11&288&1008&1546&1824& 1952 & 2008\\
12&466&1854&2980&3586 & 3872 & 4000\\
13&754&3410&5744&7050 & 7680 & 7968\\
14&1220&6272&11072&13860 & 15234 &15872\\
15&1974&11536&21342&27248& 30218 & 31616\\
\hline
\end{tabular}
}
\caption{Sequences $b_n^{(k)}$ for some fixed values of $k$.} \label{bn}
\end{center}
\end{table}

\bigskip

Obviously the set $B_n^{(0)}(0^k,1^k)$ coincides with $Z_n^{(k)}(0^k,1^k)$, hence

\begin{equation}\label{zn}
z_n^{(k)}=\left \{
\begin{array}{ll}
1 & \mbox{if} \ n =0\\
\\
b_n^{(k)}/\ 2& \mbox{if} \ n \geq 1.
\end{array}
\right.
\end{equation}

%\subsection{The sequence $z_n^{(k)}$}
%
%By a suitable inspection, it is not difficult to note that the construction of $Z_n(0^k,1^k)$ is the same of the one of $R_n(0^k,1^k)$: the constraint of the initial $1$ in a string in $R_n(0^k,1^k)$ do not affect the argument to generate the strings in $Z_n(0^k,1^k)$. Therefore, we can immediately provide the following relation (which is the equivalent to relation (\ref{rn_meni}), clearly equipped with different initial conditions:
%
%\begin{equation}\label{zn_meni}
%z_n^{(k)}=\sum_{j=1}^{k}z_{n-j}^{(k)}-\sum_{j=k+1}^{2k-1}z_{n-
%j}^{(k)}\ \ \mbox{for}\ n\geq k+1,
%\end{equation}
%with $z_{-i}^{(k)}=0$, $z_0^{(k)}=1$, $z_j^{(k)}=2^{j-1}$ for $j=1,2,\ldots k-1$, and $z_k^{(k)}=2^{k-1}-1$.
%It can be proved by induction, with a similar way to the proof of Proposition \ref{rn_positive}, that:
%
%\begin{equation}\label{zn_positive}
%z_n^{(k)}=
%\displaystyle\sum_{j=1}^{k-1} z_{n-j}^{(k)}\ \ \mbox{for}\ n\geq k+1,
%\end{equation}
%with the same initial conditions.
%
%

\subsection{Generating functions and $k$-generalized Fibonacci numbers relation}

The well-known $k$-generalized Fibonacci numbers
$\{f_n^{(k)}\}_{n\geq 0}$, can be defined as
\begin{equation}\label{kbonacci}
f_n^{(k)}=\left \{
\begin{array}{ll}
2^n & \mbox{if} \ 0 \leq n \leq k-1\\
\\
f_{n-1}^{(k)}+f_{n-2}^{(k)}+\ldots+
f_{n-k}^{(k)} & \mbox{if} \ n \geq k\ .
\end{array}\right.
\end{equation}

We recall that $f_n^{(k)}$ is the number of length $n$ binary strings avoiding $0^k$. The rational generating function $f^{(k)}(x)$ of the sequence $\left\{f_n^{(k)}\right\}_{n \geq 0}$ (see A000045 for $k=2$, A000073 for $k=3$, A000078 for $k=4$ in The On-line Encyclopedia of Integer Sequence) is given by
\begin{equation}\label{knacci}
f^{(k)}(x)=\frac{1+\displaystyle\sum_{i=1}^{k-1}x^i}{1-\displaystyle\sum_{i=1}^{k}x^i} \ .
\end{equation}

Since the sequences (\ref{nobordi}) and (\ref{kbonacci}) are very similar, it is worthwhile to find a link between the two sequences. It is not difficult to show, by induction, that
\begin{equation}\label{relation1}
b_n^{(k)}=\left \{
\begin{array}{ll}
1 & \mbox{if} \ n = 0\\
\\
2f_{n-1}^{(k-1)} & \mbox{if} \ n \geq 1.
\end{array}
\right.
\end{equation}

Therefore,
\begin{equation}\label{relation_z}
z_n^{(k)}=\left \{
\begin{array}{ll}
1 & \mbox{if} \ n = 0\\
\\
f_{n-1}^{(k-1)} & \mbox{if} \ n \geq 1.
\end{array}
\right.
\end{equation}

The rational generating functions $b^{(k)}(x)$ and $z^{(k)}(x)$ for the sequence (\ref{relation1}) and (\ref{relation_z}), respectively, are

$$
b^{(k)}(x)= 2x f^{(k-1)}(x) +1
$$
and
$$
z^{(k)}(x)=xf^{(k-1)}(x) +1\ .
$$

Similarly to the sequences $b_n^{(k)}$ and $z_n^{(k)}$, also $r_n^{(k)}$ can be expressed in terms of $k$-generalized Fibonacci numbers as showed in the following proposition.

\begin{proposition}\label{prop_bordi}
\begin{equation}\label{bordi}
r_n^{(k)}=\left \{
\begin{array}{ll}
1 & \mbox{if} \ n = 0\\
\\
\frac{f^{(k-1)}_{n-1} + d^{(k)}_n}{2} & \mbox{if} \ n \geq 1,
\end{array}
\right.
\end{equation}
\end{proposition}

\emph{Proof.} \quad We can proceed by induction. If $n=1$, then $r_1^{(k)}=(f_0^{k-1}+d_1^{(k)})/2=0$ which is the same value obtained by recurrence (\ref{rn_meni}). Suppose that $r_s^{(k)}=\frac{f^{(k-1)}_{s-1} + d^{(k)}_s}{2}$ for each $s<n$. We have, from Proposition \ref{rn_positive},
$$r_n^{(k)}=r_{n-1}^{(k)}+\ldots+r_{n-k+1}^{(k)} +d^{(k)}_n \ .$$
For inductive hypothesis
\begin{equation*}
\begin{array}{rl}
r_n^{(k)}&=\frac{f^{(k-1)}_{n-2} + d^{(k)}_{n-1}}{2}+ \ldots + \frac{f^{(k-1)}_{n-k} + d^{(k)}_{n-k+1}}{2} +d_n^{(k)}\\\\

&=\frac{f^{(k-1)}_{n-2}+\ldots+f^{(k-1)}_{n-k}}{2}+\frac{2d_n^{(k)}+d_{n-1}^{(k)}+\ldots+d_{n-k+1}^{(k)}}{2}\\\\

&=\frac{f_{n-1}^{(k-1)}+d_n^{(k)}}{2}+\frac{d_n^{(k)}+d_{n-1}^{(k)}+\ldots+d_{n-k+1}^{(k)}}{2}.
\end{array}
\end{equation*}

Since
$\left\{d_n^{(k)}\right\}_{n\geq 0}=\Big\{1,-1,\underbrace{0,0,\ldots,0}_{k-2},1,-1,\underbrace{0,0,\ldots,0}_{k-2},1,-1,0,\ldots\Big\}$ and
$\sum_{i=0}^{k-1}d_{n+i}^{(k)}=0$ for each $n\geq 0$, then the thesis follows. \cvd

The rational generating function $d^{(k)}(x)$ of the sequence $\left\{d_n^{(k)}\right\}_{n\geq 0}$
(see sequences A049347 for $k=3$, A219977 for $k=4$, in The On-line Encyclopedia of Integer Sequence) is given by
$$
d^{(k)}(x)= \frac{1}{1+\displaystyle\sum_{i=1}^{k-1}x^i} \ ,
$$

and, from Proposition \ref{prop_bordi}, the generating function $r^{(k)}(x)$ of the sequence $\left\{r_n^{(k)}\right\}_{n\geq 0}$ can be easily obtained as
$$
r^{(k)}(x)= \frac{x f^{(k-1)}(x) + d^{(k)}(x)-1}{2}\ .
$$

Following the previous arguments, using (\ref{relation1}), (\ref{relation_z}) and (\ref{bordi}), recalling that $3\leq k\leq \left\lfloor\frac{n}{2}\right\rfloor$, $m \geq 2$, $n \geq 2k$, and noting that $b_0^{(k)}=1$ and $r_2^{(k)}=1$, the cardinality of $S_{m \times n}^{(k)}$ given in (\ref{totmatrix}) can be computed in terms of $k$-generalized Fibonacci numbers and $d^{(k)}_n$ as

\begin{equation*}
|\mathcal S_{m \times n}^{(k)}|=\left \{
\begin{array}{ll}

\left(
f_{2k-1}^{(k-1)}
\right)^{m-2}& \mbox{if} \ n = 2k\\

& \\

f_{n-2k-1}^{(k-1)} \cdot
 \left(f_{n-2k+1}^{(k-1)}+d_{n-2k+2}^{(k)}\right) \cdot
 \left(f_{n-1}^{(k-1)} \right)^{m-2} &
 \mbox{if} \ n > 2k \ .
\end{array}
\right.
\end{equation*}

%r_n^{(k)}=\left \{
%\begin{array}{ll}
%1 & \mbox{if} \ n = 0\\
%\\
%\frac{f^{(k-1)}_{n-1} + d^{(k)}_n}{2} & \mbox{if} \ n \geq 1,
%\end{array}
%\right.

Note that if $k=3$ it is possible to derive a closed formula for (\ref{totmatrix}) using the following well-know closed formula for Fibonacci numbers (adapted to our purposes)

$$
f^{(2)}_n= \frac{1}{\sqrt{5}}\left( \left(\frac{1+\sqrt{5}}{2}\right)^{n+2} - \left(\frac{1-\sqrt{5}}{2}\right)^{n+2}\right) \  \mbox{for} \ n \geq 0 \ .
$$

For $k>3$ it is possible to observe that  the generating function for the sequence $\left\{\mathcal S_{m \times n}^{(k)}\right\}_{n>2k}$ is rational due to the fact that the Hadamard product of rational generating functions is rational (see \cite{lando}).  We recall that the \textit{Hadamard product} of two generating functions
$$
A(x)=a_0+a_1x+a_2x^2+\ldots\quad \mbox{and}\quad
B(x)=b_0+b_1x+b_2x^2+\ldots
$$

\noindent
is the generating function
$$
(A\circ B)(x)=a_0b_0+a_1b_1x+a_2b_2x^2+\ldots\ .
$$

\section{Conclusions and further developments}

A set of non-overlapping strings is often referred as a non-overlapping code or cross-bifix-free code. We refer to \cite{black} for an exhaustive list of references on the subject.
The properties of string codes have been object of deep research in the last fifty years (see for example \cite{berst}).
Extending the definition of a string code, a
two dimensional code is defined (see \cite{ans2}) as a set $X$ of matrices over a finite alphabet $\Sigma$ if any matrix over $\Sigma$ has at most one tiling decomposition with elements of $X$. In this sense, the set
$\mathcal S_{m\times n}^{(k)}$ can be seen as a two-dimensional code. It should be worth to study the properties of $\mathcal S_{m\times n}^{(k)}$ and to investigate if they have similarities with the usual properties of string codes.

\medskip

As well as in the linear case, an interesting line of 
research could take into consideration the construction 
of a non-overlapping set $N$ of matrices having fixed 
dimension $m \times n$ which is also non-expandable, that 
is, for each $A \in U \backslash N$ there exists a matrix 
$B \in N$ such that $A$ and $B$ are overlapping matrices. 
%(otherwise, it could be interesting to prove that the 
%non-expandable property is lost because of the 
%generalization to the two-dimensional case)
It is not difficult to see that a matrix $A \in \mathcal S^{(k)}_{m\times n}$ and a matrix $B \in \mathcal S^{(k')}_{m\times n}$ are overlapping matrices, for each $k,k'$ with $k \neq k'$ and $3\leq k, k'\leq \left\lfloor\frac{n}{2}\right\rfloor$.

Actually, there exist matrices in $U$ which can be added in $\mathcal S^{(k)}_{m\times n}$ in order to keep the non-overlapping property. An example is constituted by the set of the matrices of the form showed in Figure \ref{ganascia} which are non-overlapping with the matrices in $\mathcal S^{(3)}_{6\times 10}$.
\begin{figure}
$$
\begin{pmatrix}\
\bf{1}&\bf{1}&\bf{1}&\bf{0}&*&*&*&\bf{1}&\bf{0}&\bf{0}\\
*&*&*&*&*&*&*&*&*&\bf{0}\\
*&*&*&*&*&*&*&*&*&\bf{0}\\
*&*&*&*&*&*&*&*&*&\bf{0}\\
*&*&*&*&*&*&*&*&*&\bf{0}\\
\bf{1}&\bf{1}&\bf{1}&\bf{1}&*&*&*&\bf{0}&\bf{0}&\bf{0}
\end{pmatrix}
$$
\caption{The structure of the matrices expanding $\mathcal S^{(3)}_{6\times 10}$.}
\label{ganascia}
\end{figure}
Such matrices are obtained by adding at the beginning of the first and last row an entry 1 in the matrices in $\mathcal S^{(3)}_{6\times 10}$ and deleting an entry of
that rows chosen among the ones with index $j$, with $k <  j < n-k$. We can extend this process obtaining the following family of sets.

\begin{definition}\label{matrixZ}
Let $3\leq k\leq \left\lfloor\frac{n}{2}\right\rfloor$ and let $0 \leq h \leq n-2k$.
We denote $\mathcal S^{(k,h)}_{m\times n}\subset \mathcal M_{m\times n}$ the set of the matrices $A=\left(a_{i,j}\right)$ satisfying the following conditions:

\begin{itemize}
		
\item $A_1=1^h1^{k-1}0w_110^{k-1}$, where $v_1=0w_11$ is a binary string of length $n-2k+2-h$ avoiding both $0^k$ and $1^k$;

\item for $i=2,\ldots,m-1$, $A_i=w_i0=v_i$, where $v_i$ is a binary string of length $n$ avoiding both $0^k$ and $1^k$;

\item $A_m=1^h 1^kv_m0^k$, where $v_m$ is a binary string of length $n-2k-h$ avoiding both $0^k$ and $1^k$.
		
\end{itemize}
\end{definition}

Note that, in the case $h=0$, the set $\mathcal S^{(k,h)}_{m\times n}$ coincides with $\mathcal S^{(k)}_{m\times n}$. The following proposition can be easily proved.

\begin{proposition} The set
$\mathcal S^{(k,h)}_{m\times n} \cup \mathcal S^{(k,h+1)}_{m\times n}$
is non-overlapping, for each $k$ with $3\leq k\leq \left\lfloor\frac{n}{2}\right\rfloor$, $0 \leq h <n-2k$, $m \geq 2$ and $n \geq 2k$.
\end{proposition}

On the contrary, the set $\mathcal S^{(k,h)}_{m\times n} \cup \mathcal S^{(k,h+j)}_{m\times n}$, with $j \geq 2$, does not maintain the non-overlapping property since an overlap is possible by means of a slipping of the last row of a matrix $A \in \mathcal S^{(k,h)}_{m\times n}$ on the first row of a matrix $B \in \mathcal S^{(k,h+j)}_{m\times n}$.

\medskip

Clearly, the matrices in the set we have defined have fixed dimensions $m\times n$. If matrices of different dimensions are considered in the same set, the definition of non-overlapping matrices should be slightly revised moving from Definition 2.1 and taking care to the case when one matrix is a proper submatrix of another one. In this direction, a
further analysis could examine under which conditions on $m,n$
and $k$ the set $\displaystyle\bigcup_{m,n,k}\mathcal
S_{m\times n}^{(k)}$ is still a non-overlapping set or a two
dimensional code. For this purpose, a first easy result should
be the following: fixed $k$ and $n$, the set $\mathcal
S_{m\times n}^{(k)}\cup \mathcal S_{m'\times n}^{(k)}$, with
$m<m'$, is not non-overlapping since each matrix
$A\in \mathcal S_{m\times n}^{(k)}$ can be completely
overlapped on a matrix $B\in \mathcal S_{m'\times n}^{(k)}$
having the block partition
$B=\begin{bmatrix}
B_1\\
A
\end{bmatrix}
$.


\begin{thebibliography}{1}

\bibitem{ans} M. Anselmo, D. Giammarresi, M. Madonia.
Two-Dimensional Rational Automata: A Bridge Unifying One- and Two-Dimensional Language Theory
\emph{Lecture Notes in Computer Science} {\bfseries 7741} 133--145, 2013.

\bibitem{ans2} M. Anselmo, D. Giammarresi, M. Madonia.
Structure and Measure of a Decidable Class of Two-dimensional Codes. \emph{Lecture Notes in Computer Science} {\bfseries 8977} 315--327, 2015.

\bibitem{ans3} M. Anselmo, D. Giammarresi, M. Madonia. Unbordered Pictures: Properties and Construction. \emph{Lecture Notes in Computer Science} {\bfseries 9270} 45--57, 2015.

\bibitem{tcs} E. Barcucci, A. Bernini, S. Bilotta, R. Pinzani. Cross-bifix-free sets in two dimensions. \emph{Theoretical Computer Science}. In Press (2015). \texttt{http://dx.doi.org/10.1016/j.tcs.2015.08.032}

\bibitem{bajic} D. Bajic, T. Loncar-Turukalo. A simple suboptimal construction of cross-bifix-free codes. \emph{Cryptography and
Communications} {\bfseries6} 27--37, 2014.

\bibitem{berst} J. Berstel, D. Perrin, C. Reutenauer. Codes and automata (Encyclo-
pedia of Mathematics and its Applications). Cambridge University Press,
Cambridge, 2009.


\bibitem{bilo} S. Bilotta, E. Pergola, R. Pinzani.
A new approach to cross-bifix-free sets.
\emph{IEEE Transactions on Information Theory} {\bfseries58} 4058--4063, 2012.


\bibitem{black} S. Blackburn. Non-overlapping codes. \emph{IEEE Transactions on Information Theory} {\bfseries61} 4890--4894, 2015.

\bibitem{singa} Y. M. Chee, H. M. Kiah, P. Purkayastha, C. Wang.
Cross-bifix-free codes within a constant factor of optimality.
\emph{IEEE Transactions on Information Theory} {\bfseries59} 4668--4674, 2013.

\bibitem{croc} M. Crochemore, C. Iliopoulos, M. Korda. Two-dimensional prefix string matching and covering on square
matrices. \emph{Algorithmica} {\bfseries 20} 353--373, 1998.

%\bibitem{gilbert} E.N. Gilbert. Synchronization of binary messages. \emph{IRE Transactions on Information Theory}
%{\bfseries 6} 470–477, 1960.

\bibitem{kita}  S. Kitaev, T. Mansour, A. Vella.
Pattern Avoidance in Matrices.
\emph{Journal of Integer Sequences} {\bfseries 8} Article 05.2.2, 2005.

\bibitem{lando} S. K. Lando. Lecture on Generating Functions. American Mathematical Society, 2003.

\bibitem{nielsen} P. T. Nielsen. A Note on Bifix-Free Sequences.
\emph{IEEE Transactions on Information Theory}, vol. IT-29,
704-706, September 1973.

\end{thebibliography}
\end{document}